\documentstyle[prb,aps, multicol]{revtex}

\def\etal{\emph{et al.}}

\input BoxedEPS
\SetepsfEPSFSpecial
\HideDisplacementBoxes

\begin{document}

\title{Magnetoresistance of single-domain ferromagnetic particles}
\author{J.\ Aumentado and V.\ Chandrasekhar}
\address{Department of Physics and Astronomy, Northwestern University, Evanston, IL
60208} 
\maketitle

\begin{abstract} We have performed magnetoresistance measurements on single-domain, 
submicron elliptical Ni particles using nonmagnetic probes in a four probe geometry
at liquid helium temperatures.  In the smallest particles, the magnetoresistance shows sharp jumps which are
associated with the switching of individual domains.  Using an anisotropic magnetoresistance model, we can
reconstruct hysteresis loops of the normalized magnetization.  The remanent magnetization in
zero applied magnetic field is typically 15 percent less than the saturation magnetization.  This
relaxation of the magnetization may be due to surface effects or crystal grain structure in the particles.
\end{abstract}


\begin{multicols}{2}

The study of micromagnetics is currently experiencing a revival as experimental
techniques improve to the point where it is possible to observe the magnetization
dynamics of single submicron-sized particles.
While the methods for studying single particle micromagnets have so far
been limited to magnetoforce microscopy (MFM),\cite{mfm} Lorentz microscopy\cite{lorentz} and microbridge
SQUID  magnetometry,\cite{wern95,wern}
transport-based micromagnetic measurements may provide a robust alternative which is viable
over a large temperature scale. 
Until recently, experimental efforts at studying electrical transport in mesoscale magnets have
mostly concentrated on arrays of thin wires,\cite{shearwood,adeyeye} arrays of chains of 
particles,\cite{martin} and very long, thin single wires.\cite{hg95}  However, future
mesoscopic spin-transport devices are likely to incorporate single-domain ferromagnetic particles in
combination with non-magnetic elements.  Consequently, it is important that the transport characteristics
of single-domain particles be well understood.

In this Letter, we report on our measurements of the magnetoresistance of patterned
submicron single-domain elliptical Ni thin films in magnetic fields in the plane of the films.  At low
fields, a hysteretic magnetoresistance is observed, similar to that seen by Hong and Giordano in long
Ni wires.\cite{hg95,hg97}  In addition, a single discrete jump or a series
of discrete jumps can also be observed which are associated with magnetization switching in the domains of the
particle.  Using an anisotropic
magnetoresistance (AMR) model,\cite{mcpotter} we can deduce the normalized magnetization of the particles
as a function of magnetic field.  The resulting hysteresis loops show that the remanent magnetization is
smaller than the saturation magnetization, which may be due to surface effects or
misalignment of the direction of magnetization in the crystal grains of the particle.  

Figure 1 shows scanning electron micrographs of one of our typical samples, along with a schematic
illustrating the device structure.  The samples were fabricated onto an oxidized Si substrate using
conventional bilayer e-beam lithography and liftoff techniques in two steps. In the first step, the elliptical
particles were  patterned and deposited so that they lay flat on the substrate.  Ni (99.995\% pure, 30 nm
thick) was deposited in an e-gun 
\begin{figure}[p]
\begin{center}
\BoxedEPSF{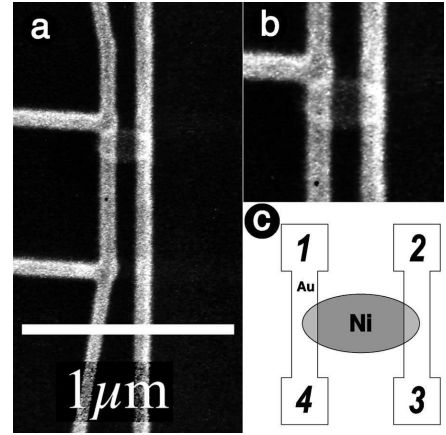 scaled 750}
\end{center}
\caption{SEM and schematic representations of devices: (a)  typical probe and magnet
configuration, (b) enlargement of Ni  ellipse,  (c) schematic  of four probe
measurement.}
\label{devSEM}
\end{figure}
evaporator at a  base pressure of $5\times10^{-8}$ Torr at a rate of 0.01
nm/s, resulting in films with a resistivity of approximately 8.3 $\mu\Omega$-cm at
4.2 K, as measured on a coevaporated  1~$\mu$m wide Ni thin film.  The elliptical geometry constrained the
easy axis in-plane along the major axis of the ellipse.  The series of particles discussed here had lateral
dimensions of either 120~nm$\times$240~nm or 220~nm$\times$640~nm. This size was chosen because particles of
similar dimensions measured by microbridge SQUID magnetometry  have demonstrated sharp, single
magnetization switches and nominally square hysteresis loops in fields applied along
the easy axis.\cite{wern95}  In the second lithography step, Au probes were patterned 
to make electrical contact to the particle.  Prior to
the Au deposition, the exposed Ni contact areas were cleaned using an ac Ar$^{+}$ etch. Au (99.999\% pure,
60 nm thick) was then deposited by e-gun evaporation,
creating two narrow (70~nm  width) nonmagnetic probe wires extending across the
ellipse along the minor axis direction.
Four terminal measurements using these probes allowed us to measure the resistance of the ferromagnetic
particle, along with a small contribution from the magnetic/nonmagnetic interface.  

The use of nonmagnetic
probes is a unique and essential component of our sample design.  The only other transport experiments in
particles of this size range of which we are aware used probes fabricated from the same material as the
sample.\cite{jia97}  This configuration is problematic since the magnetic probes will modify the
magnetization of the sample. As the probe size becomes comparable to the sample size, this effect will become
more pronounced.  In addition, the injection of spin-polarized carriers from the magnetic probes into the
magnetic particle might also be expected to affect the magnetoresistance.\cite{singleelement}  We avoid these
difficulties by eliminating the ferromagnetic probes and using nonmagnetic Au probes to make contact to our devices.

The samples were measured in a pumped $^4$He
cryostat with the major axis of the ellipses aligned parallel or perpendicular (to within $5^{\circ}$) to
the applied magnetic field direction. The resistance was measured in a four-probe
configuration using standard ac lock-in/bridge techniques with excitation currents
between 0.1 and  5.0 $\mu$A at 11.7 Hz at temperatures between 1.2 and 10~K. The
traces were identical at all excitation currents in this range, with the exception 
of an improved signal-to-noise ratio at higher currents. In this probe
configuration the direction of the current is primarily along the major axis of the ellipse.  
\begin{figure}[p]
\begin{center}
\BoxedEPSF{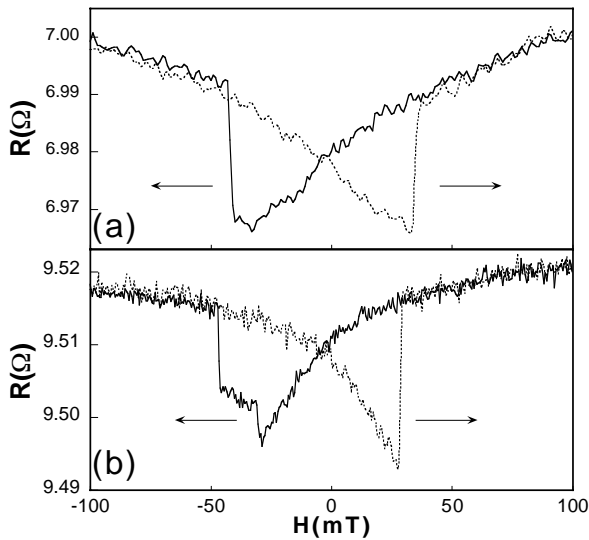 scaled 900}
\end{center}
\caption{Parallel field magnetoresistance traces of Ni ellipses at T=1.5~K.  
(a) 120 nm $\times$  240~nm Ni ellipse. (b) 220~nm$\times$640~nm ellipse.}
\label{MRs}
\end{figure}
Figure 2(a) shows a typical low-field parallel magnetoresistance trace taken at
$T =$1.5~K in a 120~nm$\times$240~nm elliptical particle. 
The magnetic field is swept between $\pm2000$ G at a constant rate of less than 2 G/s. 
Following the sweep from positive-to-negative magnetic field
(solid trace), one can see a single discrete jump in the magnetoresistance,
while the negative-to-positive sweep (dotted trace) shows similar behavior in the opposite direction. 
Before the jumps, the magnetoresistance is completely reversible. 
After the jumps,  the traces then become reversible on a second curve which is nearly mirror symmetric with
the original curve.
In other samples, the magnetization reversal may occur in multiple steps.  Figure 2(b) shows similar
traces for a 220~nm$\times$640~nm elliptical particle.  The negative-to-positive field sweep (dotted line)
shows a single jump as before, but two distinct jumps are observed in the opposite sweep direction (solid
line).   The single large jump in the negative-to-positive
sweep indicates a complete switch in the magnetization direction occuring in one step, while the two
distinct steps in the reverse direction may be due
to an intermediate magnetization state that would be found in the presence of domain structure or
magnetization pinning.  We note that small positive jumps in the magnetoresistance have also been seen in
long wires by Hong and Giordano in studies of domain wall scattering.\cite{hg97}  

These results demonstrate that one might infer the magnetic state of a
single-domain particle by measuring its magnetoresistance.  In order to do this, one needs to understand the
mechanism which gives rise to the magnetoresistance.  Hong and Giordano\cite{hg95} attributed the
magnetoresistance in their Ni wires to scattering of electrons by magnons.  However, in our samples, the
temperature independence of the magnetoresistance is inconsistent with this explanation.  
Shearwood \etal \cite{shearwood} proposed instead that the similar magnetoresistance behavior they observed
in their arrays of permalloy wires arose from the well-known anisotropic magnetoresistance (AMR) of
ferromagnets.\cite{mcpotter}
The AMR of ferromagnets is due to an anisotropic spin-orbit interaction which leads to a resistance sensitive
to the angle between the current density and the magnetization.  The resistivity due to AMR can be expressed
as\cite{mcpotter}
\begin{equation}
\rho=\rho_{\perp}+\Delta \rho_{AMR}cos^{2}\xi,
\label{eq:eqn1}
\end{equation}
where $\Delta\rho_{AMR}\equiv\rho_{\parallel}-\rho_{\perp}$ and $\xi$ is the 
angle between the current and magnetization. $\rho_{\parallel}$ ($\rho_{\perp}$) is the extrapolation to zero
magnetic induction ($B=0$) of the magnetoresistance with the magnetic field parallel (perpendicular) to
the direction of the current.  Experimentally, $\Delta\rho_{AMR}$ is found to be positive, so that the
resistance with the magnetization parallel to the current direction is greater than the resistance with the
magnetization perpendicular to the current direction.  We believe that the magnetoresistance of our particles
arises from this effect.

Consider again the magnetoresistance of the single small ferromagnetic particle shown in Fig. 2 in
terms of the AMR model.  At large negative magnetic fields, the magnetization is saturated in
the direction of the field, and the resistance has its highest value, consistent with the AMR model.  As the
field is increased, i.e., the magntitude of the field is decreased, the resistance of the sample
decreases.  If we make the assumption that the magnetization 
throughout the ellipse is uniform
and rotates coherently relative to the current direction, this decrease in resistance indicates that the
direction of the  magnetization is changing as a function of field, since the direction and magnitude of the
current are constant.  We can express the projection of the magnetization along the current direction (or
major axis of the ellipse) as
$M_{x}=M_{0}cos\xi$ (where $M_{0}$ is the saturation magnetization). From Eq. (\ref{eq:eqn1}), the normalized
magnetization
$m_x(H) = M_x /M_0$ can be expressed in terms of the field dependent resistance $\rho (H)$ as
\begin{equation} |m_{x} (H)|=\left(\frac{\rho (H)-\rho_{\perp}}{\Delta\rho_{AMR}}\right)^{\frac{1}{2}}.
\label{eq:eqn2}
\end{equation}
Using this equation, hysteresis loops of the normalized magnetization projection may
be reconstructed from the resistance, inserting appropriate signs at the switches.
\begin{figure}[p]
\begin{center}
\BoxedEPSF{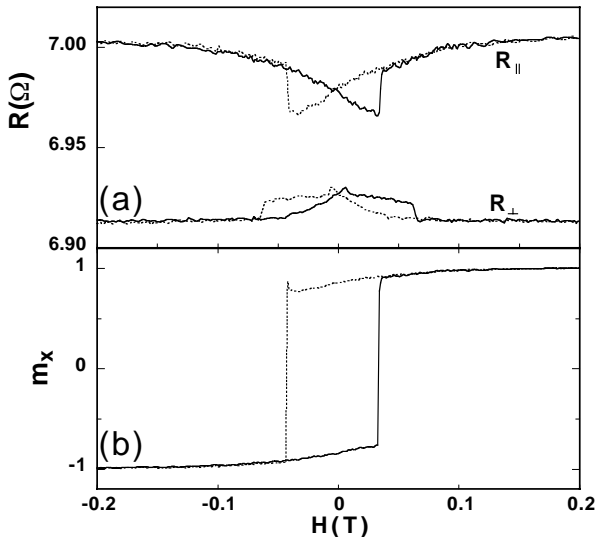 scaled 900}
\end{center}
\caption{Magnetization reconstruction from magnetoresistance of 120 nm $\times$  240
nm (30 nm thick) Ni ellipse at T=1.5~K.  (a) parallel and perpendicular field
magnetoresistances (in-plane), (b)  reconstructed normalized magnetization, $m_{x}$,
from the parallel  field magnetoresistance. The solid curve is taken while sweeping
the  field in the positive direction. The dotted curve is taken in the  negative
direction.}
\label{reconstruction}
\end{figure}
Figure 3(a) shows the magnetoresistance of the single Ni particle of Fig. 2(a) in magnetic fields both
perpendicular and parallel to the major axis of the ellipse, with the current along the major axis in both
cases.  From these measurements, one can obtain the AMR ratio, 
$\Delta\rho_{AMR}/\rho_{ave}$ (where $\rho_{ave}\equiv(\rho_{\parallel}+2\rho_\perp)/3$),
which varies between 1.25 and 1.5\% for our samples. This is somewhat lower than results in bulk 
Ni\cite{mcpotter} but is consistent with expectations in specimens of reduced
dimensionality.\cite{rijks}  Figure 3(b) shows the normalized magnetization obtained from the data of
Fig. 3(a) as described above.  The magnetization shows clean switches at applied magnetic fields of
$\sim \pm 350$ G, indicative of the presence of a single domain whose magnetization switches direction at
these fields.  Prior to the switch, however, the magnetization decreases to values less than the saturation
magnetization at low fields, dropping to approximately 0.85 of the saturation magnetization at zero
field.  In the spirit of the coherent rotation model, this would correspond to a misalignment of the
magnetization from the major axis of the ellipse of $\sim 20-30^{\circ}$ in the samples we have studied. 
More generally, a remanence ratio of less than unity might indicate the presence of a nonuniform magnetization in the
sample in zero applied field. 

What are the possible origins of a nonuniform magnetization in our particles?  A number of mechanisms can be
suggested.  First, the particles are elliptical films, and not ellipsoids of revolution which
means that the magnetization in
the particle is not uniform.  In high magnetic fields, all magnetic moments in the particle would be aligned along the major
axis of the ellipse, but at low fields, the moments at the surface of the film in particular would tend to be
canted due to the nonuniform demagnetizing field.  A rough order of magnitude estimate of this misalignment
can be obtained by considering the moments in a rim of width of the thickness of the film around the
perimeter of the ellipse to be completely perpendicular to the major axis of the ellipse.  With the
dimensions of the sample of Fig. 3, this gives an estimate of 0.75 for the remanence ratio, in good
agreement with the value of 0.85 we observe.  Second, a remanence ratio of less than unity may be
due to the polycrystalline nature of the Ni films.  If the small Ni particle consists of a few crystal grains
with randomly oriented crystalline axes, the magnetization of each grain in zero applied field may relax along
crystal axes due to crystalline anisotropy.  This would lead to reduction in the component of the total
magnetization parallel to the major axis of the ellipse.  Finally, modification of the spin-orbit interaction
close to the surface is also expected to lead to the misalignment of moments near the surface.\cite{aharoni} 
At present, we cannot distinguish between these different mechanisms, but we are inclined towards the
nonuniform demagnetizing field as the simplest mechanism\cite{shearwood} which would account for our observations.  We
should note that remanence ratios approaching unity have been observed in magnetization measurements on similarly
sized particles by Wernsdorfer \etal \cite{wern95} using microbridge SQUIDs, but we believe that these
results might be affected by flux exclusion due to the Meissner effect in the superconducting microbridge
devices.  There are no nearby magnetic elements in our devices which might affect the magnetic
environment of the Ni particles.

Future one-dimensional spin-transport
devices will typically incorporate small ferromagnetic elements, and it is
important to understand the resistance of these elements before the overall operation of such
devices can be understood.  In addition to their importance to device physics, transport measurements on
ferromagnetic films provide a means of studying the magnetic properties of single domain particles over wide
ranges in field and temperature which are not accessible by other measurement techniques. 

The authors wish to acknowledge extensive discussions with Anupam Garg,  John
Ketterson, and Yuli Lyanda-Geller. This work was supported by the David and Lucile
Packard Foundation and  the MRSEC program of the National Science Foundation
(DMR-9632472) at  the Materials Research Center of Northwestern University.

\end{multicols}


\begin{references}

\bibitem{mfm}M. Lederman \etal , J. Appl. Phys. {\bf 73}, 6961 (1993); M. Lederman \etal , J. Appl. Phys.
{\bf 75}, 6217 (1994); M. Lederman \etal , Phys.
Rev. Lett. {\bf 73}, 1986 (1994); R. O'Barr \etal , J. Appl. Phys. {\bf 79}, 5303 (1996).

\bibitem{lorentz}C. Salling \etal , J. Appl. Phys. {\bf 75}, 7989 (1994).

\bibitem{wern95}W. Wernsdorfer \etal , J. Magn. Magn. Mater. {\bf 145}, 133 (1995).

\bibitem{wern}W. Wernsdorfer \etal , Phys. Rev. Lett. {\bf 77}, 1873 (1996); W. Wernsdorfer \etal , Phys. Rev.
B {\bf 55}, 11552 (1997); W. Wernsdorfer \etal , Phys. Rev. Lett. {\bf 78}, 1971 (1997); W. Wernsdorfer \etal
, Phys. Rev. Lett. {\bf 79}, 4014 (1997); W.T. Coffey \etal , Phys. Rev. Lett. {\bf 80}, 5655 (1998).

\bibitem{shearwood}C. Shearwood \etal , J. Appl. Phys. {\bf 75}, 5249 (1994).

\bibitem{adeyeye}A.O. Adeyeye \etal , Appl. Phys. Lett. {\bf 70}, 1046 (1997).

\bibitem{martin}J.I. Martin \etal , Appl. Phys. Lett. {\bf 72}, 255 (1998).

\bibitem{hg95}K. Hong and N. Giordano, Phys. Rev. B {\bf 51}, 9855 (1995).

\bibitem{hg97}K. Hong and N. Giordano, J. Phys. C. (1998).

\bibitem{mcpotter}T.R. McGuire and R.I. Potter, IEEE Trans. Mag. {\bf MAG-11}, 1018 (1975).

\bibitem{jia97}Y.Q. Jia \etal , J. Appl. Phys. {\bf 81}, 5461 (1997).

\bibitem{singleelement}M. Viret \etal , Phys. Rev. B {\bf 53}, 8464 (1996); J.F. Gregg \etal , Phys. Rev. Lett. {\bf
77}, 1580 (1996).

\bibitem{rijks}T.G.S. Rijks \etal , Phys. Rev. B {\bf 51}, 283 (1995).

\bibitem{aharoni}A. Aharoni, \emph{Introduction to the Theory of Ferromagnetism} [Oxford University Press,
New York (1996)].

\end{references}
\end{document}